\documentclass[aps,pre,superscriptaddress,reprint]{revtex4-2} 

\usepackage{amsmath}
\usepackage{amssymb}
\usepackage{esint}
\usepackage{graphicx}
\usepackage{dcolumn}
\usepackage{bm}
\usepackage[separate-uncertainty,per-mode=symbol]{siunitx}

\graphicspath{{figures/}}

\renewcommand{\vec}[1]{\boldsymbol{#1}}
\renewcommand{\tensor}[1]{\underline{#1}}
\newcommand{\abs}[1]{\left\vert #1 \right\vert}
\newcommand{\avg}[1]{\left< #1 \right>}
\newcommand{\real}[1]{\Re\left\{ #1 \right\}}
\newcommand{\order}[1]{\mathcal{O}\left\{ #1 \right\}}

\newcommand{\fixme}[1]{{\color{red}[[#1]]}}

\newcommand{\hypgeo}[2]{%
  \operatorname{%
    {\vphantom{\mathnormal{F}}}_{#1}%
    \kern-\scriptspace
    \mathnormal{F}_{#2}%
  }%
}

\begin{document}

\title{Scattered waves fuel emergent activity}

\author{Ella M. King*}
\affiliation{Department of Physics and Center for Soft Matter Research,
New York University, New York, NY 10003, USA}
\affiliation{Simons Junior Fellow,
 160 5th Ave, New York, NY 10010, USA}

\author{Mia C. Morrell*}
\affiliation{Department of Physics and Center for Soft Matter Research,
New York University, New York, NY 10003, USA}

\author{Jacqueline B. Sustiel}
\affiliation{Department of Physics and Center for Soft Matter Research,
New York University, New York, NY 10003, USA}

\author{Matthew Gronert}
\affiliation{Department of Physics and Center for Soft Matter Research,
New York University, New York, NY 10003, USA}

\author{Hayden Pastor}
\affiliation{Department of Physics and Center for Soft Matter Research,
New York University, New York, NY 10003, USA}

\author{David G. Grier}
\affiliation{Department of Physics and Center for Soft Matter Research,
New York University, New York, NY 10003, USA}

\date{\today}

\begin{abstract}
Active matter taps into external
energy sources to power its own processes.
Systems of passive particles ordinarily lack this capacity, but can become active 
if the constituent particles interact
with each other nonreciprocally.
By reformulating the theory of classical
wave-matter interactions,
we demonstrate that interactions mediated by scattered waves
generally are not constrained by Newton's third law.
The resulting center-of-mass forces
propel clusters of scatterers, enabling them to extract
energy from the wave and rendering them
active.
This form of activity is an emergent property of the scatterers' state of
organization and can arise in any system where mobile objects scatter waves.
Emergent activity flips the script on
conventional active matter
whose nonreciprocity emerges
from its activity, and not the other way around.
We combine theory, experiment and
simulation to illustrate how emergent activity
arises in wave-matter composite systems and
to explore the phenomenology of emergent activity
in experimentally accessible models.
These preliminary studies suggest that heterogeneity
is a singular perturbation to the dynamics of
wave-matter composite systems, and induces
emergent activity under all but the most
limited circumstances.
\end{abstract}

\maketitle

\section{Introduction}
\label{sec:introduction}

Active matter harvests energy from its environment
and reuses that energy for its own purposes,
for example to power its own
motion \cite{ramaswamy2010mechanics,marchetti2013hydrodynamics,bechinger2016active, shankar2022topological}.
Familiar examples include the molecular motors 
within cells \cite{needleman2017active},
synthetic colloidal swimmers 
\cite{palacci2013living},
bacterial swarms \cite{darnton2010dynamics},
and flocks of birds \cite{cavagna2014bird}.
The individual entities
that make up such systems are
``active particles'' in the sense that they
independently transduce energy.

\begin{figure}
    \centering
    \includegraphics[width=\columnwidth]{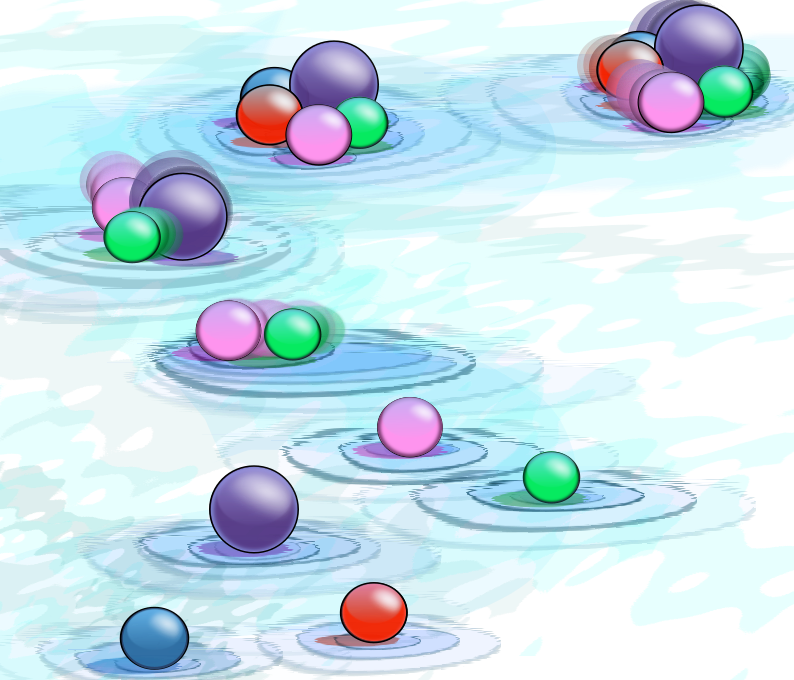}
    \caption{Depiction of emergent activity in a wave-matter
    composite system. Individual particles scatter the wave symmetrically
    and so experience no net force. Scattered waves mediate interactions
    that coalesce groups of particles into pairs, trios, and rafts. Unless the particles have identical scattering properties, the net wave-mediated
    interaction transfers energy from wave into the clusters' translational
    and rotational kinetic energy.}
    \label{fig:intro}
\end{figure}

An alternative form of activity can arise when otherwise 
passive particles interact through
nonreciprocal forces.
The imbalance in these forces
enables groups of particles to translate and rotate even when the individual particles
have no mechanism to acquire the
necessary energy.
Hallmarks of this form of activity have been observed in 
simulations of two-component interdiffusing fluids~\cite{you2020nonreciprocity}, 
dispersions of catalytically active particles~\cite{soto2014self}, 
and robotic metamaterials~\cite{brandenbourger2019non}, 
all of which display transitions between passive and active states
enabled by nonreciprocal interactions.
These disparate systems exemplify a 
broad category of compositions 
of matter that display ``emergent activity''
in the sense that their activity is an
emergent property of their
state of organization.
Whereas previous studies have focused on simulations of
specialized model systems,
we demonstrate
that emergent activity arises ubiquitously in any context
where mobile objects scatter waves.
More specifically, we demonstrate, perhaps counterintuitively, that the wave-mediated interactions that bind scatterers
into clusters need not be reciprocal.
Figure~\ref{fig:intro}
illustrates how the interplay of
reciprocal and nonreciprocal
wave-mediated interactions can
organize independent passive scatterers
into active clusters.

We develop the principle of emergent activity in Sec.~\ref{sec:emergentactivity}
by focusing on sound and light as archetypal waves.
In both cases, we derive analytic expressions for
the wave-mediated interactions between dissimilar particles, 
and identify conditions under which the pair interaction is
nonreciprocal.
Experimental observations in Sec.~\ref{sec:experiments}
and simulations in Sec.~\ref{sec:simulations} reduce the theory to practice, illustrating
how unbalanced acoustic forces give rise to
emergent activity in monolayers
of acoustically levitated granular matter.
These studies reveal that heterogeneity in
the particles' scattering properties 
constitutes a singular perturbation to 
the system's dynamics, which means that
emergent activity should be a generic feature
of any system composed of particles that
scatter waves.
We conclude in Sec.~\ref{sec:discussion}
with a discussion of the general consequences of 
emergent activity for self-organization in natural and 
synthetic many-body systems.

\section{The principle of emergent activity}
\label{sec:emergentactivity}

Active particles are self-propelled and so can
move autonomously through their environment.
Passive particles, by contrast, lack the ability
to transduce energy on their own.
They therefore
relax into equilibrium configurations unless driven by external forces.
Emergent activity provides a mechanism for
certain configurations of
passive particles to develop the capacity to move autonomously.
This kind of collective self-propulsion arises
in systems of passive particles that
interact with each other through nonreciprocal forces.

\begin{figure}
    \centering
    \includegraphics[width=0.6\columnwidth]
    {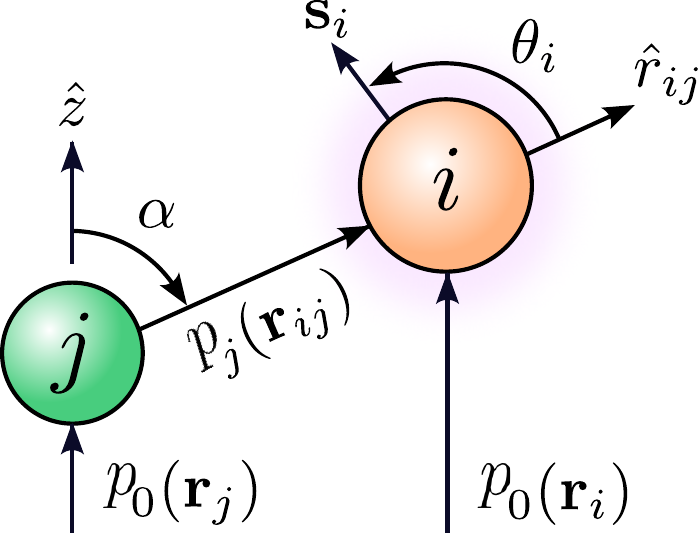}
    \caption{Geometry for computing sound-mediated forces and interactions.
    Particle $j$ scatters a portion of the incident pressure
    wave, $p_0(\vec{r})$, to
    its neighbor at $\vec{r}_i$.
    The scattered wave, $p_j(\vec{r}_{ij})$, 
    contributes to the force experienced by particle $i$.
    Formulating this influence is facilitated
    by defining a spherical coordinate
    system centered on particle $i$ and aligned
    with the separation between the particles, $\vec{r}_{ij} = \vec{r}_i - \vec{r}_j$.}
    \label{fig:geometry}
\end{figure}

The two-particle system depicted in Fig.~\ref{fig:geometry} illustrates the
physical basis for emergent activity
and helps to clarify the distinction between 
conventional activity, simple driving and
emergent activity.
A passive particle located at $\vec{r}_i$ experiences
a force, $\vec{F}_{ij}(\vec{r}_{ij})$, due to
its interaction with its neighbor at $\vec{r}_j$.
This force depends on the 
interparticle separation,
$\vec{r}_{ij} = \vec{r}_i - \vec{r}_j$, as well as the two particles' properties.
The same coupling mechanism
also mediates a force, $\vec{F}_{ji}(\vec{r}_{ji})$,
on particle $j$ that is
obtained from $\vec{F}_{ij}(\vec{r}_{ij})$
by exchanging the particles' labels.
Newton's third law leads us to expect these forces
to be reciprocal in the sense that
$\vec{F}_{ij}(\vec{r}_{ij}) =  -\vec{F}_{ji}(\vec{r}_{ji})$.
If the interparticle interaction is nonreciprocal,
however, then the net force,
\begin{equation}
    \Delta\vec{F}_{ij}(\vec{r}_{ij})
    =
    \vec{F}_{ij}(\vec{r}_{ij})
    +
    \vec{F}_{ji}(\vec{r}_{ji}) ,
\end{equation}
 acting on the
pair's center of mass need not vanish.

To show how such nonreciprocity can arise, we refer 
again to Fig.~\ref{fig:geometry} and
consider what happens when passive particles
interact with an incident wave,
$p_0(\vec{r}, t)$, that acts as a reservoir of
energy, momentum and angular momentum.
In scattering the incident wave, 
each particle experiences a primary time-averaged force,
$\vec{F}_i(\vec{r}_i)$,
that could be used to drive its motion.
Instead, we assume that the particles come to
mechanical equilibrium
on manifolds where $\vec{F}_i(\vec{r}_i) = 0$.
In that sense, the wave does not drive the particles.

In addition to the primary force,
particle $i$ also experiences 
a secondary force, $\vec{F}_{ij}(\vec{r}_{ij})$, 
due to the wave, $p_j(\vec{r}_{ij}, t)$,
that is scattered by its neighbor.
Particle $j$ experiences an analogous secondary
force mediated by $p_i(\vec{r}_{ji}, t)$.
In the next two Sections, we
formulate the time-averaged
secondary interactions, $\vec{F}_{ij}(\vec{r}_{ij})$,
for particles scattering either sound or light
and demonstrate that those interactions are nonreciprocal
unless the particles have identical scattering properties.
Analytical expressions for
the net center-of-mass force,
$\Delta\vec{F}_{ij}(\vec{r}_{ij})$,
reveal how scattering enables pairs
of passive particles to propel themselves
autonomously through the wave.
The pair of scatterers
therefore is active even though the individual particles are not.

For clarity, we focus on the time-averaged forces
exerted by harmonic waves \cite{bruus2012acoustofluidics}.
We restrict our attention to
spherical particles whose scattering patterns
are readily expressed as multipole expansions.
We furthermore work in the Rayleigh limit,
considering particles that are substantially
smaller than the wavelength, $\lambda$, of
the incident wave, and invoke the first Born approximation, limiting
the analysis to first-order scattering.
The interaction then is conveniently expressed in
orders of the dimensionless size parameters,
$k a_i$ and $k a_j$, where $a_i$ and $a_j$ are
the radii of the two spheres and $k = 2 \pi/\lambda$
is the wavenumber.
These simplifying assumptions are not
required for wave-mediated interactions to
display nonreciprocity.
They are useful because they
clarify the origin and nature of 
nonreciprocal interactions in model systems that
lend themselves to detailed analysis.

\subsection{Nonreciprocal interactions mediated by sound}
\label{sec:nonrec_sound}

The force landscape created by a general acoustic wave 
includes both nonconservative driving forces and 
conservative trapping forces \cite{abdelaziz2020acoustokinetics}.
To avoid confusion between driving and activity,
we specialize to the case of a planar standing wave that exerts no
driving forces.
A pair of particles levitated in such a standing wave
experiences a pressure field,
\begin{equation}
    p_0(\vec{r})
    =
    p_0 \sin(k z),
\end{equation}
that we take to be aligned in the vertical direction, $\hat{z}$.
As discussed in Appendix~\ref{sec:acousticforces},
a particle that is denser and less compressible
than the medium experience a primary scattering
force, $\vec{F}_i(\vec{r}_i)$, that localizes it at a node
of the pressure field, where the primary force vanishes.
In formulating the interaction between a pair of levitated
spheres, we assume that both
are localized in the same nodal plane at $z = 0$.
Referring to Fig.~\ref{fig:geometry}, this
corresponds to scattering angle $\alpha = \pi/2$.
\begin{subequations}
    \label{eq:konig}
The corresponding scattering-mediated interaction between  particles $i$
and $j$ is derived in Appendix~\ref{sec:particles} 
and has the leading-order form,
\begin{equation}
    \label{eq:konigforce}
    \vec{F}^K_{ij}(\vec{r}_{ij})
    =
    - 
    2 \pi \, F_0 \,
    \Phi(kr_{ij}) \,
    \eta_i \eta_j \,
    \hat{r}_{ij},
\end{equation}
which historically has been called
the K\"onig interaction \cite{konig1893hydrodynamisch}.
The overall scale of the K\"onig interaction
is set by the intensity of the incident wave,
\begin{equation}
    \label{eq:F0}
    F_0
    =
    \frac{p_0^2}{\rho_0 \omega^2} .
\end{equation}
Its dependence on particle separation,
\begin{equation}
    \Phi(x)
    =
    \frac{1}{x^4} 
    \left[
    \left(1 - \frac{x^2}{3}\right) \cos(x)
    + x \sin(x)
    \right] ,
\end{equation}
is shaped by the scattered wave's interference
with the incident wave.
The particles' properties enter into
$\vec{F}^K_{ij}(\vec{r}_{ij})$ through
dimensionless coupling constants of the form
\begin{equation}
    \eta_i
    =
    \frac{\rho_0 - \rho_i}{\rho_0 + 2 \rho_i} 
    (k a_i)^3 ,
\end{equation}
each of which depends on the mismatch between the 
particle's density, $\rho_i$, and that of the medium, $\rho_0$,
and is proportional to the particle's volume.
Equation~\eqref{eq:konig} includes all 
contributions up to $\order{(ka_i)^3 (ka_j)^3}$ in the reduced particle
sizes and is reciprocal under exchange of the labels $i$ and $j$.
Surprisingly, the expression in Eq.~\eqref{eq:konig} for the leading-order K\"onig
interaction between dissimilar
spheres appears not to have been reported previously.

The K\"onig interaction is attractive for small separations \cite{weiser1984red,silva2014acoustic}
and so tends to draw monolayers of levitated
particles together into close-packed rafts
\cite{kobelev1979self,apfel1999studies,rabaud2011acoustically,abdelaziz2021ultrasonic,lim2019cluster,lim2022mechanical}.
The pair interaction is a central force because
the two-particle system is symmetric
and the wave itself carries
no angular momentum \cite{bliokh2019spin}.

\end{subequations}
Whereas the leading-order K\"onig interaction is reciprocal,
the multipole expansion in Appendix~\ref{sec:particles}
reveals that next-order contributions to the
interparticle force break that symmetry:
\begin{subequations}
\label{eq:acousticforce}
\begin{equation}
    \vec{F}_{ij}(\vec{r}_{ij})
    =
    (1 + \chi_{ij}^K) \,
    \vec{F}^K_{ij}(\vec{r}_{ij}) ,
\end{equation}
where the additional factor,
\begin{equation}
    \chi_{ij}^K = 
    \alpha_{ij} 
    + \beta_{ij} (k a_j)^2
    + \gamma_{ij} (k a_i)^2,
\end{equation}
\end{subequations}
includes all contributions up to
$\order{(ka_i)^5}$ and $\order{(ka_j)^5}$.
The coefficients of $\chi_{ij}^K$ depend
on the densities and compressibilities
of the two particles and are
reported in Eq.~\eqref{eq:chiKij}.
These corrections
to the leading-order
K\"onig interaction are
nonreciprocal unless the two particles
have identical properties.
A pair of acoustically
levitated spheres therefore experiences a center-of-mass force,
\begin{equation}
    \label{eq:cmforce}
    \Delta\vec{F}_{ij}^K(\vec{r}_{ij})
    =
    (\chi_{ij}^K - \chi_{ji}^K) \, \vec{F}^K_{ij}(\vec{r}_{ij}),
\end{equation}
that the individual spheres would
not have felt.


The analogous formulation of the wave-mediated
pair interaction between bubbles is presented
in Appendix~\ref{sec:bubbles}.
Nonreciprocal effects tend to be much weaker for
spheres that are less dense and more compressible
than the medium.

\begin{figure*}
    \centering
\includegraphics[width=0.9\textwidth]{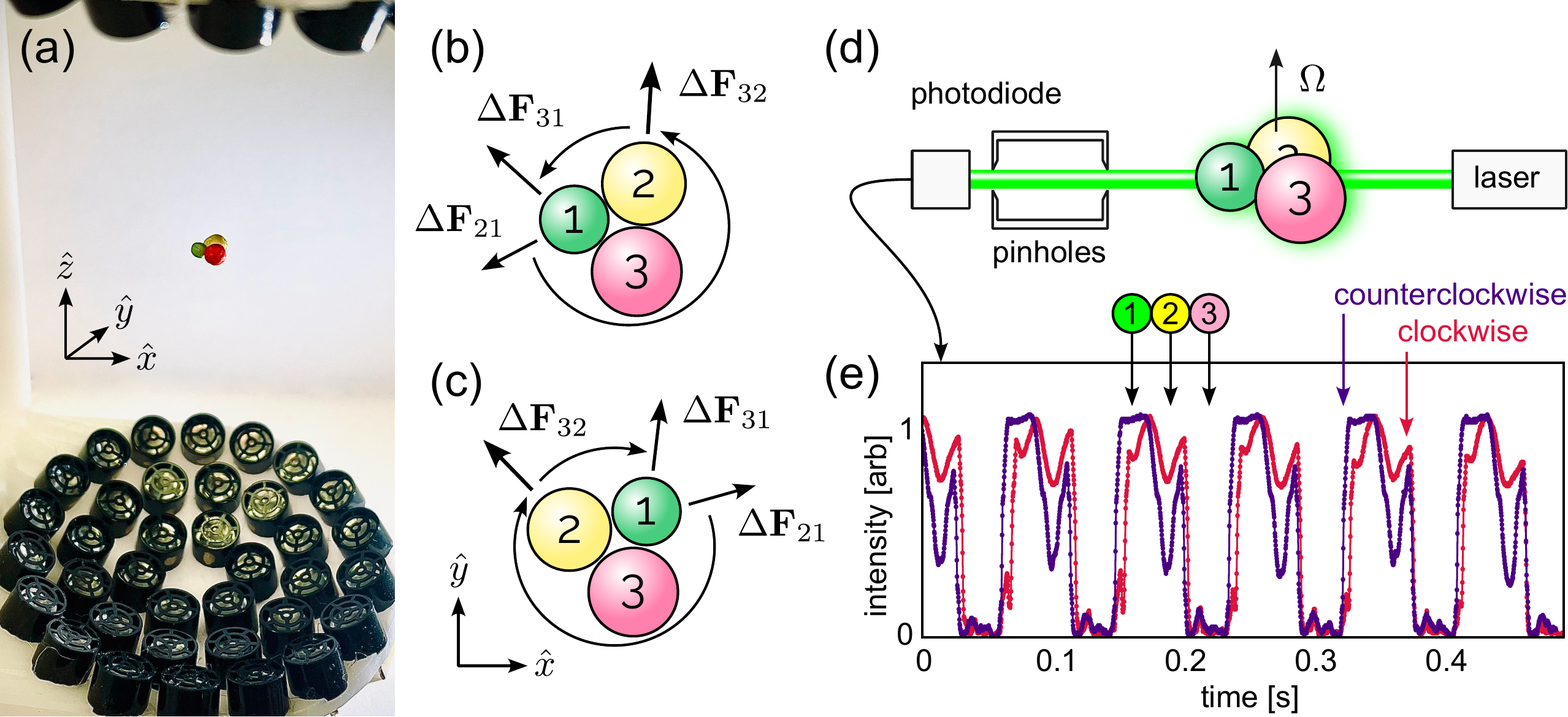}
    \caption{Emergent activity displayed by three expanded polystyrene (EPS) particles levitated in an acoustic trap. (a) Photograph of the experimental system. Three millimeter-scale particles are trapped in a single node of a standing-wave
    acoustic trap at \SI{40}{\kilo\hertz}.
    (b) Schematic representation of the unbalanced forces acting on the three pairs of particles due to nonreciprocal wave-mediated
    interactions. The resultant torque causes the cluster to rotate about its center of mass.
    The cluster also would translate were it not held in place
    by the trap.
    (c) Reversing the chirality of the cluster
    reverses the direction of rotation.
    (d) Light-obscuration measurement of the cluster's rotation.
    A collimated laser beam is partially blocked each time a particle passes through the optical axis.
    The transmitted light is filtered by a pair of pinholes
    and its intensity is recorded with a photodiode.
    (e) Typical time traces of the recorded intensity for
    the two chiral configurations show the three particles
    moving through the beam in the sequence
    predicted in (b) and (c). The same rotation rate, $\Omega
    = \SI{11(1)}{\hertz}$,
    is recorded for both rotation directions.}
    \label{fig:trio}
\end{figure*}

\subsection{Nonreciprocal interactions mediated by light}
\label{sec:nonrec_light}

Pair interactions can
be mediated by scattered electromagnetic waves
in a phenomenon known as
optical binding \cite{burns1989optical,dholakia2010colloquium}.
In this case, the role played by the
pressure wave, $p_0(\vec{r})$, in Fig.~\ref{fig:geometry}
is played instead by the light's electric field,
$\vec{E}_0(\vec{r})$.
Transverse optical binding occurs when
when the interparticle separation is
perpendicular to the light's axis of propagation,
$\alpha = \pi/2$.
This interaction recently has been formulated
for metallic nanoparticles \cite{sukhov2015actio}
and has been shown to be nonreciprocal
if the particles have different
dipole polarizabilities.
Here, we report the complementary
result for a pair of dielectric spheres
when $\vec{E}_0(\vec{r})$ is linearly
polarized perpendicularly to the
interparticle separation, $\vec{r}_{ij}$.
This form of transverse optical binding
is analogous to the acoustic K\"onig interaction.
Appendix~\ref{sec:opticalbinding}
adapts the Green function formalism
of Ref.~[\onlinecite{dapasse1994optical}]
to obtain the leading-order optical binding force,
\begin{subequations}
\label{eq:opticalbinding}
\begin{equation}\label{eq:PhiO}
    \vec{F}_{ij}(\vec{r}_{ij})
    =
     -\frac{3}{2} F_0 \real{\Phi(kr_{ij}) \, \alpha_i \alpha_j^*}\,
    \,
    \hat{r}_{ij},
\end{equation}
whose overall scale,
\begin{equation}
    F_0 = k \abs{\vec{E}_{0}}^2,
\end{equation}
is proportional to the light's intensity.
As in the acoustic case, the dependence 
on particle separation,
\begin{equation}
    \Phi(x)
    = 
    \left( -\frac{1}{3}x^2 + i x + 1 \right)
    e^{ix} ,
\end{equation}
is structured by interference between the
scattered wave and the incident wave.
Assuming again that the particles are smaller than the wavelength, their leading-order coupling to the field is set by
dimensionless dipole polarizabilities
\cite{draine1993beyond,albaladejo2010radiative},
\begin{equation}
\label{eq:polarizability}
    \alpha_i
    =
    \frac{\alpha^{(0)}_i}{
    1 - i \frac{\alpha^{(0)}_i}{6 \pi \epsilon_0 n_0^2}},
\end{equation}
where $\epsilon_0$ is the permittivity of space,
$n_0$ is the refractive index of the medium and
\begin{equation}
\label{eq:clausiusmossotti}
    \alpha^{(0)}_i
    =
    4 \pi \epsilon_0 \, n_0^2 \, 
    \frac{n_i^2 - n_0^2}{n_i^2 + 2 n_0^2} \, (k a_i)^3
\end{equation}
\end{subequations}
is the Clausius-Mossotti polarizability
for a sphere of radius $a_i$ and
refractive index $n_i$.
Equations~\eqref{eq:polarizability} 
and \eqref{eq:clausiusmossotti}
are suitable
for dielectric particles in the Rayleigh limit, $k a_i \ll 1$.
In that limit, the leading contribution
to the light-mediated
interaction is proportional to $(k a_i)^3 (k a_j)^3$ 
and is reciprocal under exchange of the particles' labels \cite{dapasse1994optical}.

Retaining contributions to next order
in the reduced size parameters,
$k a_i$ and $k a_j$,
yields an expression,
\begin{subequations}
    \label{eq:opticalforce}
\begin{equation}
    \vec{F}_{ij}(\vec{r}_{ij})
    =
    \real{(1 + i\chi_{ij}^O) \,
    \vec{F}^O_{ij}(\vec{r}_{ij})},
\end{equation}
that
can be separated into 
reciprocal and nonreciprocal components,
where the real part of
\begin{equation}
\label{eq:decomposedopticalforce}
    \vec{F}^O_{ij}(\vec{r}_{ij})
    =
     -\frac{3}{2} F_0 \, 
    \Phi(kr_{ij}) \,
    \left( \alpha_i' \alpha_j' + \alpha_i'' \alpha_j''\right) \,
    \hat{r}_{ij}
\end{equation}
agrees with the standard expression for
the transverse optical binding force
\cite{dapasse1994optical}.
Single and double primes 
in Eq.~\eqref{eq:decomposedopticalforce}
refer to the real and imaginary parts
of the polarizabilities, respectively.
The leading nonreciprocal corrections
are given by
\begin{equation}
    \chi_{ij}^O 
    = 
    \frac{\alpha_i' \alpha_j'' - \alpha_j' \alpha_i''}{\alpha_i' \alpha_j' + \alpha_i'' \alpha_j''}.
\end{equation}
For dielectric particles in the Rayleigh limit, the leading-order nonreciprocal
corrections,
\begin{equation}
    \chi_{ij}^O
    =
    \frac{2}{3}  \frac{n_j^2 - n_0^2}{n_j^2 + 2n_0^2}
    (k a_j)^3
    - 
    \frac{2}{3} \frac{n_i^2 - n_0^2}{n_i^2 + 2n_0^2}
    (k a_i)^3 ,
\end{equation}
\end{subequations}
depend on the particles' sizes and refractive indexes and vanish appropriately if the particles
are identical.
The observation that 
$\chi_{ij}^{O} \ne \chi_{ji}^{O}$ confirms
that light-mediated interactions generally
are nonreciprocal.
This complements the analogous result
for metallic nanospheres reported in Ref.~[\onlinecite{sukhov2015actio}].
As for the acoustic interactions discussed
in Sec.~\ref{sec:nonrec_sound}, the
expression in Eq.~\eqref{eq:opticalforce} appears
not to have been reported previously
and predicts behavior that should
be observable experimentally.
Most significantly, Eq.~\eqref{eq:opticalforce} and the
predictions of Ref.~[\onlinecite{sukhov2015actio}] together demonstrate
that emergent activity should arise naturally
in systems of particles that scatter light.

\section{Experimental observations
of emergent activity}
\label{sec:experiments}

Figure~\ref{fig:trio}(a) depicts a simple
experimental demonstration of emergent activity
in a trio of acoustically levitated spheres.
The system consists of three millimeter-scale
beads of expanded polystyrene (EPS)
\cite{horvath1994expanded} with
a measured mass density \cite{morrell2023acoustodynamic}  of
$\rho_j = \SI{30.5(2)}{\kg\per\cubic\meter}$
levitated
in a standing-wave acoustic trap.
The levitator is
based on the standard TinyLev design \cite{marzo2017tinylev} and
consists of two
banks of piezoelectric transducers
(MA40S4S, Murata, Inc.)
operating at \SI{40}{\kilo\hertz}.
Each bank of \num{36} transducers
is driven harmonically at
$\SI{12}{\volt_\text{pp}}$
by a software-defined function generator
(Arduino Teensy 4.0)
and projects a traveling wave
into a spherical volume of air
\SI{12.5}{\cm} in diameter.
Interference between the counterpropagating
waves creates a standing wave 
with pressure nodes
along the instrument's vertical axis,
$\hat{z}$.

The focused acoustic trap is more
highly structured than the plane standing wave
used to develop the theory of nonreciprocal
wave-mediated interactions in Appendix~\ref{sec:interactions}.
Nevertheless, the experimental system shares
essential features with the
idealized model.
An individual bead experiences
one of the TinyLev's pressure nodes
as a three-dimensional Hookean potential
energy well
\cite{marzo2017tinylev}
with a measured \cite{morrell2023acoustodynamic}
stiffness of \SI{5}{\micro\newton\per\mm\per\volt}
for a \SI{1.5}{\mm}-diameter EPS bead.
The well associated with one node
is large enough to contain at least three
such particles.
The trio in Fig.~\ref{fig:trio}(a) is
held in contact by a combination of the trap's
primary restoring force and
the beads' secondary wave-mediated interaction.
The balance of forces keeps the
cluster of particles rigidly trapped in the
instrument's $x$-$y$ plane even when the
instrument is inclined relative to
gravity.
Interference among incident and scattered waves 
gives rise to interparticle forces that should at least
qualitatively resemble
predictions of Eq.~\eqref{eq:acousticforce}.

We approximate the three-particle interaction 
by the superposition of pairwise forces depicted
in Fig.~\ref{fig:trio}(b).
Each of these nonreciprocal contributions
is directed along one of the pair separation
vectors, and therefore contributes to a
torque around the
cluster's center.
Of these, the unbalanced interaction
between the largest and smallest beads,
$\Delta \vec{F}_{31}$,
is predicted by Eq.~\eqref{eq:acousticforce} to be
weaker than the combined influence of the
other two contributions.
As a result, the cluster should experience
a net torque in the $\hat{z}$ direction
that causes it to rotate counterclockwise, as drawn.
Indeed, the cluster is observed to rotate rapidly
within its acoustic trap in the anticipated direction.

Exchanging any two beads, as illustrated in
Fig.~\ref{fig:trio}(c), reverses the cluster's chirality
and therefore should reverse its direction of rotation.
This also is observed in the experimental system.
A typical realization of this experiment
is presented in Supplementary Video 1.

We measure the cluster's rotation rate using
the light obscuration system depicted in
Fig.~\ref{fig:trio}(d). This eliminates the possibility of temporal
aliasing in camera measurements due to the cluster's rapid rotation.
The collimated beam from a \SI{2}{\milli\watt} modular diode laser is aligned so that
it is at least partially occluded when one of the
particles rotates into the beam.
The beam has a diameter of \SI{2}{\mm}, which is
comparable to the diameters of the particles.
Particles of different sizes can be distinguished
by the proportion of the beam they block.
The transmitted light passes through two
coaxial \SI{250}{\um}-diameter pinhole apertures
separated by \SI{20}{\mm} before being recorded
by a photodiode. 
The photocurrent is
digitized with a storage oscilloscope
(TDS2002, Tektronix)
at \SI{5000}{samples\per\second}.

Typical time traces of the recorded laser intensity
are plotted in Fig.~\ref{fig:trio}(e)
and confirm that the sense of rotation
places the smallest particle
in the lead, followed by the mid-sized
particle and then the largest.
For the specific trio of particles
captured in Fig.~\ref{fig:trio}(a),
the measured rotation rate is
$\Omega = \SI{11(1)}{\hertz}$.
The cluster rotates at the same rate in
either chiral configuration, which confirms
that the torque results from the particles'
configuration and is not somehow encoded into
the structure of the acoustic trap.
The cluster's rotation therefore is an example
of emergent activity.


\section{Emergent activity in simulated many-body systems}
\label{sec:simulations}

To gain insight into the bulk behavior of emergently active matter,
we simulate large ensembles of acoustically levitated particles.
We first simulate a trio of dense spheres to confirm
that the pairwise approximation at least qualitatively
accounts for the experimental observations in
Sec.~\ref{sec:experiments}.
We then explore larger systems to understand
how nonreciprocal forces affect many-body dynamics.

\begin{figure}
    \centering
    \includegraphics[width=0.75\columnwidth]{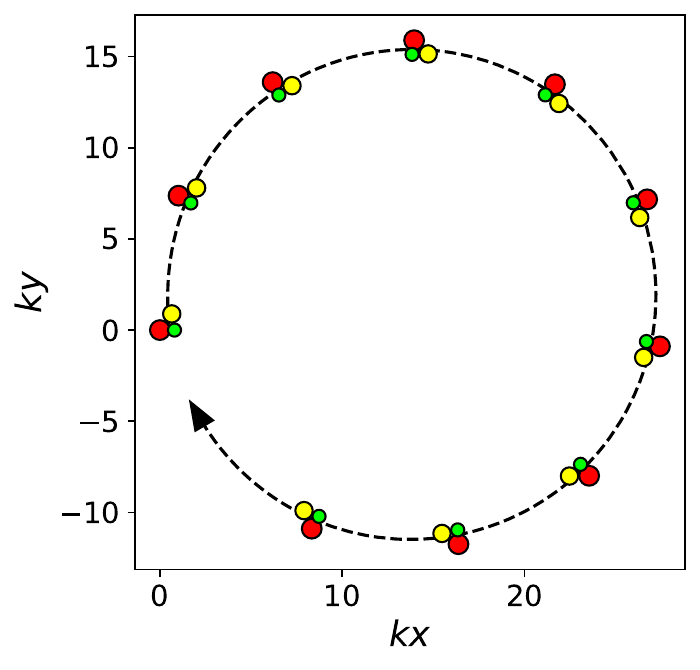}
    \caption{Simulated trajectory of a three-particle cluster of dense spheres,
    translating as it rotates in the nodal plane of an acoustic standing wave. The particles'
    sizes and density are chosen to resemble
    the experimental system in Sec.~\ref{sec:experiments}, as is the
    force scale, $F_0$.}
    \label{fig:spinny}
\end{figure}

\begin{figure*}
    \centering
    \includegraphics[width=\textwidth]{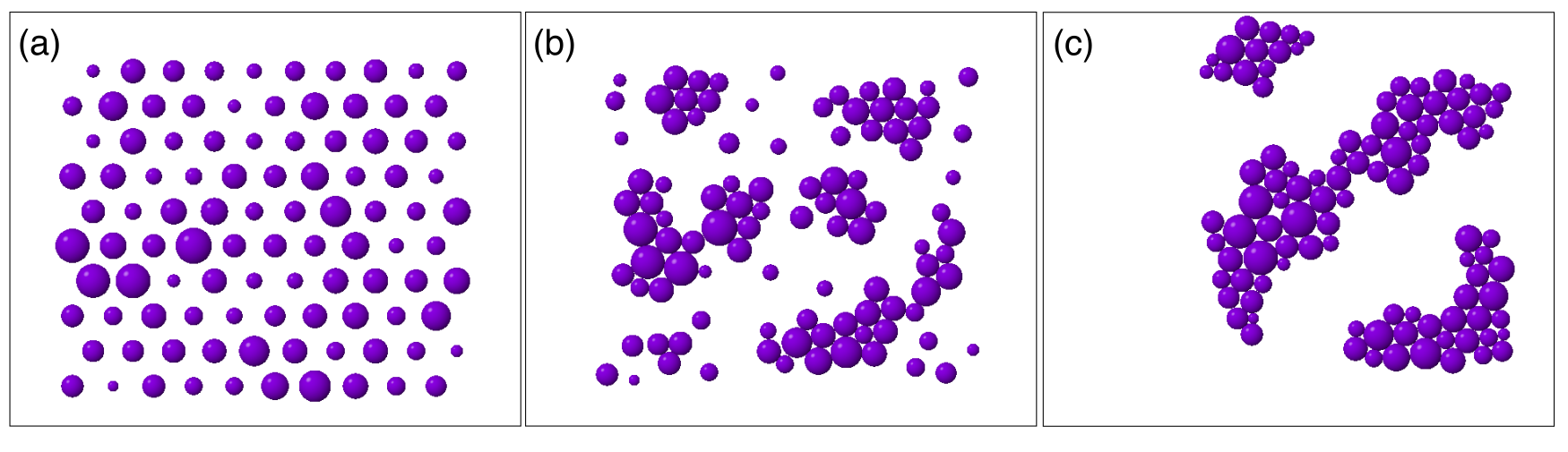}
    \caption{ Snapshots from a simulation of 
    \num{100} acoustically levitated spheres with a polydispersity of $X = \sigma/\mu = \num{0.25}$.
    The mean radius $k\mu = \num{0.4}$ and
    force scale, $F_0$, are chosen to reflect
    experimental conditions from Sec.~\ref{sec:experiments}.
    (a) Particles are randomly initialized in a triangular lattice. The simulation then evolves over \num{e7} time steps with a step size of \SI{e-3}{\second}.
    (b) Clusters rapidly coalesce under the
    influence of the wave-mediated K\"onig
    interaction.
    (c) Clusters display emergent activity
    by translating, rotating, colliding and
    internally restructuring.
    }
    \label{fig:traj}
\end{figure*}

We simulate the dynamics of dense spheres
stably levitated in an acoustic plane wave
using the analytic expression for wave-mediated pair interactions,
$\vec{F}_{ij}^K(\vec{r}_{ij})$, from Eq.~\eqref{eq:acousticforce}.
Each particle moves in the plane
according to the equation of motion
\begin{equation}
    m_i \ddot{r}_i
    =
    - 
    \gamma_i \, \dot{r}_i
    + 
    \sum_{j \neq i} \vec{F}^K_{ij}(\vec{r}_{ij}) + \frac{\varepsilon}{\sigma_{ij}} \left(1 - \frac{r_{ij}}{\sigma_{ij}}\right)^{\alpha - 1},
\end{equation}
with a Stokes drag coefficient,
$\gamma_j = 6 \pi \eta_0 \, a_i$,
that depends on the 
viscosity of air,
$\eta_0 = \SI{1.8e-5}{\pascal\second}$ \cite{rumble2023crc},
and where the last term describes a soft-sphere steric repulsion that prevents particle overlap ($\epsilon = \num{1e5}$, $\sigma_{ij} = a_i + a_j$, and $\alpha = 4$).
The simulations employ a velocity
Verlet integrator based on the integrator
in the JAX-MD molecular dynamics
engine \cite{schoenholz2021jax}. 
We maintain accuracy up to single precision.
We set the overall force scale
to $F_0 = \SI{10}{\micro\newton}$,
which is strong enough to rigidly confine the spheres to the plane. 
This drive-to-drag ratio 
is consistent with the forces estimated \cite{morrell2023acoustodynamic} for
the experiments in Sec.~\ref{sec:experiments},
with typical K\"onig interactions of \SI{3}{\micro\newton} and Stokes drag per unit velocity of 
\SI{0.5}{\micro\newton\cdot\s\per\m}. 
Analogous simulations
of nonreciprocal optical
binding in systems of dielectric spheres interacting
through $\vec{F}^O_{ij}(\vec{r}_{ij})$ from Eq.~\eqref{eq:opticalforce}
are presented in the Supplemental Information.
The acoustic and optical systems both display emergent activity
with similar phenomenology.

\subsection{Three particles}
\label{sec:threeparticles}
To make contact with the experiments from Sec.~\ref{sec:experiments}, 
we simulate a cluster of three particles that
are composed of the same material but
differ in size, with reduced radii of
$ka_1 = \num{0.3}$, $ka_2 = \num{0.5}$,
and $ka_3 = \num{0.8}$.
For consistency with the experimental observations, 
we set the density of the particles to be \num{30} 
times that of the medium \cite{morrell2023acoustodynamic}.

Without the confining potential of the experimental acoustic trap, 
the simulated trio of particles
translates across the nodal plane as it rotates.
Figure~\ref{fig:spinny} shows a typical example,
and an animation rendered in INJAVIS~\cite{engel2021injavis} is presented in Supplementary Video 2.
The trio's coupled rotations and translations qualitatively 
resemble the meandering trajectories observed in simulations 
of larger clusters of identical spheres \cite{stclair2023dynamics}
and experiments on self-propelled bubble pairs \cite{doinikov2023self}.
Both of those systems, however, rely on viscous streaming to
break spatiotemporal symmetry.
Motion in the present case, by contrast, unambiguously emerges
from linear superposition of nonreciprocal pair interactions
mediated by scattered waves.

While the center-of-mass trajectory plotted in Fig.~\ref{fig:spinny} is circular,
other examples can be elliptical because the viscous drag on the cluster
depends on its orientation relative to the direction of motion.
Circular trajectories emerge when the rotation rate is phase locked
to the center-of-mass translation.

\subsection{Many-particle systems}
\label{sec:manyparticle}

The bulk behavior of 
emergently active systems
is tuned by the heterogeneity of the particles' properties.
Heterogeneity appears to be a singular perturbation to
the dynamics of particle pairs.
Larger ensembles may behave differently 
if they coalesce into configurations where nonreciprocal
forces cancel and activity consequently vanishes.

\begin{figure}
    \centering
    \includegraphics[width=\columnwidth]{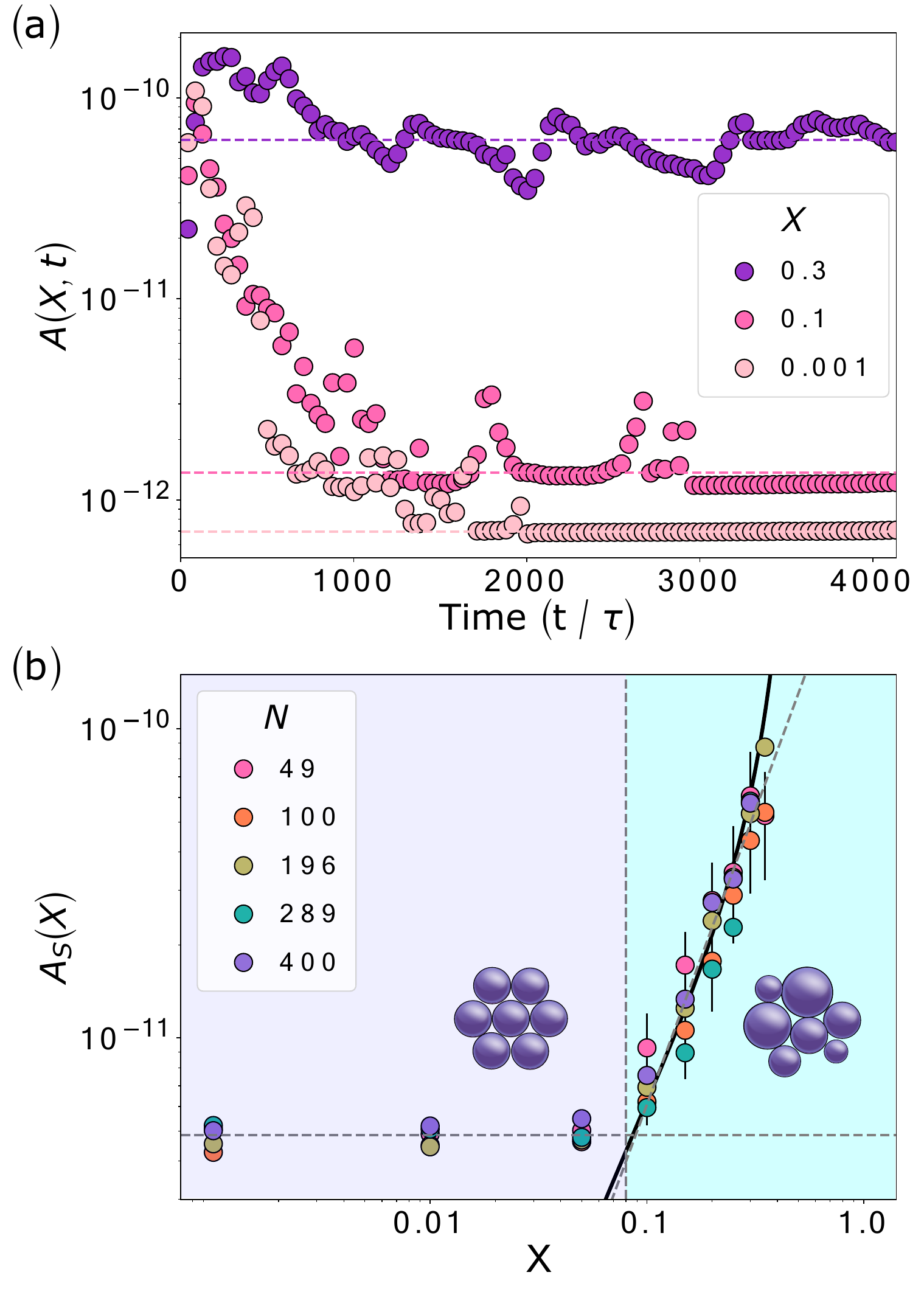}
    \caption{
    (a) Simulated time evolution of the activity, $A(X, t)$,
    of rafts of $N = \num{289}$ particles 
    of mean radius $k \mu = \num{0.4}$
    in the nodal plane of an acoustic standing wave. 
    Trajectories represent typical realizations in systems
    with varying polydispersity.
    Steady-state activities are represented by
    horizontal dashed lines.
    Dimensionless time is computed relative to the drag, $\tau = \frac{9}{2} \frac{\eta_0}{\mu^2}$.
    (b) Steady-state activity as a function of polydispersity.
    Error bars reflect repeated independent realizations.
    Polydisperse systems with $X > X^\ast$ display
    $A_S(X) \sim X^2$,
    as depicted by the gray dashed line.
    The two-particle prediction from Eq.~\eqref{eq:2particletheory} is shown as a solid black curve.
    Insets are representative configurations for $X < X^\ast$ (ordered) and for $X > X^\ast$ (disordered).
    }
    \label{fig:polydisp}
\end{figure}

We investigate collective effects in 
emergently active systems through
simulations of acoustically levitated particles 
in ensembles with particle number
ranging from $N = 49$ to $N = 400$. 
To facilitate comparison with experiments, we consider spherical particles that all have the same composition
and differ only in size.
We randomly draw the particles' radii from a normal distribution with a mean radius, $\mu$, that we
fix relative to the wavelength of sound at $k \mu = \num{0.4}$.
We vary the standard deviation of the distribution, $\sigma$, to study the effects of polydispersity.
The system 
is initialized by arranging the particles
in random order on a triangular lattice
with lattice constant $3(\mu + \sigma)$,
as shown in Fig.~\ref{fig:traj}.
The particles rapidly coalesce into clusters
under the influence of the attractive part of the
K\"onig interaction.
After this initial transient, the free-floating
clusters continue to translate and rotate
because of unbalanced nonreciprocal interactions.

One measure of such a system's activity
is provided by its rate of energy dissipation,
\begin{equation}
    P(t) = 6 \pi \eta_0 \sum_{i=1}^N a_i \, v_i^2(t),
\end{equation}
where $v_i(t)$ is the translation speed
of particle $i$ at time $t$.
This can be compared with the maximum possible dissipation
rate, 
\begin{equation}
    P_0 \equiv N F_0 v_0 ,
\end{equation}
for a system of $N$ particles all moving at terminal velocity,
\begin{equation}
\label{eq:v0}
    v_0 = \frac{F_0}{6 \pi \eta_0 \mu} ,
\end{equation}
under the influence of acoustic forces.
The result is a dimensionless activity metric,
\begin{equation}
\label{eq:activity}
    A(X, t) 
    \equiv \frac{P(t)}{P_0}
    = 
    \frac{1}{N} \sum_{i=1}^N 
    \frac{a_i}{\mu}
    \left[\frac{v_i(t)}{v_0}\right]^2,
\end{equation}
that should depend on the polydispersity, $X = \sigma/\mu$,
but should be independent of system size
for sufficiently large $N$.
Once the system reaches steady state,
this metric gauges the
rate at which the system extracts energy
from the sound wave and deposits it as
heat in the fluid medium.
Our activity metric thus tracks the rate of entropy
production by the wave-matter composite system.

Because emergent activity is a consequence
of nonreciprocal interactions, $A(X, t)$
vanishes in systems
composed of identical particles ($X = 0$).
Figure~\ref{fig:polydisp}(a) shows how the
activity evolves in time for
systems with $N = \num{289}$ particles
and three representative values of polydispersity.
In each case, the system undergoes
an initial transient from a high-activity state
as the particles coalesce into clusters.
Once clusters have formed, each system settles
into a configuration with a steady-state
activity, $A_S(X)$, that depends on its polydispersity.
Fluctuations in the long-time activity
reflect rearrangements within clusters and collisions
between clusters.
The horizontal dashed lines in Fig.~\ref{fig:polydisp}(a)
are estimates for $A_S(X)$ in each
of these realizations.
Simulations for each value of $X$
are repeated 15 times 
with different particle
ensembles and system sizes to obtain estimates for the mean steady-state
activity at each value of the polydispersity.

Figure~\ref{fig:polydisp}(b) shows how the ensemble-averaged
steady-state activity varies with polydispersity 
for acoustically-levitated particle rafts
of different system size, $N$.
As expected, $A_S(X)$
increases with increasing polydispersity
by an amount that is independent of $N$ for the
system sizes considered.
We estimate the limit of sensitivity for $A_S(X)$ by simulating systems of monodisperse particles
and find it to be
\num{e-14}.

Rafts of attractive spheres 
form triangular lattices
at low polydispersity.
The resulting six-fold
symmetry
favors cancellation of nonreciprocal forces
and therefore tends to suppress emergent activity.
Polydispersity disrupts crystalline order
above
$X^\ast \approx \num{0.08}$~\cite{pusey2009hard,pronk2004melting}
by creating topological defects
that serve as centers for unbalanced nonreciprocal forces.
The order-disorder transition therefore should enhance
emergent activity for $X > X^\ast$.
We observe this transition from a low-activity regime to a high-activity regime
in Fig.~\ref{fig:polydisp}(b), where there is a marked change in scaling at $X \approx X^\ast$.
This transition is qualitatively similar to the dynamical phase transition between passive and active states observed in random organization models \cite{Wilken2021Random}.

We model the activity of the emergently active state
by computing the ensemble-averaged activity of pairs of dissimilar particles.
Such a pair is a minimal model for the unbalanced nonreciprocal force
acting on a topological defect in a close-packed raft of particles.
The expectation value of the steady-state pair activity 
is computed in Appendix~\ref{sec:polydispersity} and is expected to scale with polydispersity as
\begin{equation}
\label{eq:leadingactivity}
    A_S(X) \sim X^2 
\end{equation}
for $X < 1$.
We compare this prediction for the ensemble-averaged pair
activity 
with the observed activity, $A_S(X)$, of
simulated particle rafts in Fig.~\ref{fig:polydisp}(b),
up to an overall multiplicative factor that is fit to the data.
To leading order,
the activity in the disordered phase scales as $X^2$.
The analytical model agrees with simulations for $X > X^\ast$.
Below $X^\ast$, 
the acoustically bound raft forms a triangular
lattice without topological defects, and
the residual activity is dominated by
incomplete cancellation of unbalanced
forces at the clusters' irregularly
shaped edges.

\section{Discussion}
\label{sec:discussion}

We have introduced the concept of emergent activity as an organizing principle.
Emergently active particles are individually passive, but become collectively active because of their nonreciprocal interactions.
Nonreciprocity is known to emerge in conventionally active systems
as a consequence of the individual particles'
activity \cite{furukawa2014activity,saha2019pairing,fruchart2021nonreciprocal,hickey2023nonreciprocal}.
Emergent activity, conversely, is a consequence of
passive particles' nonreciprocal
interactions. 

We have shown that 
nonreciprocal interactions arise naturally in
systems of particles that interact by
exchanging scattered waves.
These nonreciprocal interactions 
enable the particles to exchange energy and momentum 
with an external field, such as a sound wave or
a beam of light.
Other suitable mechanisms for inducing emergent activity 
include streaming flows around acoustically driven bubbles
\cite{ahmed2015selectively,luo2021biologically,doinikov2023self}, 
and wake-field interactions in dusty plasmas 
\cite{lisin2020plasma,usachev2009plasma,ivlev2015statistical}.

Wave-mediated interactions have the additional feature of drawing
particles together into cohesive rafts without the intercession
of other forces.
In other systems, such as dusty plasmas, that are characterized by repulsive
pair interactions, emergent activity dissipates unless an external force
maintains the density of particles.
In all of these cases, the degree of activity is enhanced by increasing
the degree of heterogeneity in the particles' properties.
For wave-mediated interactions in particular, the activity
represented by the rate of entropy production appears to be proportional
to the variance in the particle diameter.

In developing the phenomenology of emergent activity, we have
focused on systems of particles coupled to monochromatic plane standing waves.
Nonreciprocal interactions also can be fueled
by more general superpositions of
waves and configurations of particles.
Emergent activity therefore should arise ubiquitously in any system where
particles scatter waves, perhaps in combination with simple wave-mediating
driving.
These observations suggest that the collective motion powered by emergent activity was available to guide natural self-organization in the epoch before biological activity evolved
and so could have played a role in the emergence of life.

\begin{acknowledgments}
This work was supported by the National Science Foundation under Award No.~DMR-2104837. E.M.K.~acknowledges support from a Simons Foundation Junior Fellowship under Grant Number 1141499. 
We thank Marc Gershow, Jasna Brujic, and Ankit Vyas,  Mathias Casiulis, and Paul Demidov for helpful conversations.
\end{acknowledgments}

\appendix

\section{Sound-Mediated Forces}
\label{sec:acousticforces}

To formulate the forces and interactions
mediated by sound waves, we first introduce
a multipole expansion of the Reynolds stress tensor, whose normal component quantifies the force per unit surface area exerted by a sound wave on a particle.
Using this stress tensor, we first calculate the acoustic radiation force on a single
spherical particle immersed in a standing wave 
and then the acoustic interaction force between two spheres.
We use this framework to formulate
the interactions between solid particles
in Appendix~\ref{sec:particles} and 
between bubbles in Appendix~\ref{sec:bubbles}.

Our system consists of discrete
particles immersed in a harmonic
sound wave at frequency
$\omega$ whose spatial structure is described by the
pressure field, $p_0(\vec{r})$.
An analogous formulation can be provided
for objects scattering light, water ripples,
or any other harmonic wave.
The total acoustic pressure field, $p(\vec{r})$, is the superposition of
$p_0(\vec{r})$ and the waves scattered by particles in the system.
The pressure serves as the scalar potential for the sound's
velocity in a medium of density $\rho_0$,
\begin{subequations}
\label{eq:meatgrinder}
\begin{equation}
    \label{eq:velocity}
    \vec{v}(\vec{r})
    =
    - \frac{i}{\rho_0 \omega} \nabla p,
\end{equation}
in the approximation that the fluid's viscosity
may be neglected
\cite{yosioka1955acoustic,silva2011expression,abdelaziz2020acoustokinetics, doinikov1994sphere}.

Both the pressure and the velocity fields
contribute to the
time-averaged stress tensor in the fluid medium
\cite{doinikov1994sphere,sapozhnikov2013radiation},
\begin{equation}
    \label{eq:stresstensor}
    \tensor{\sigma}(\vec{r})
    =
     \frac{1}{2}
    \left[
    \kappa_0
    \abs{p(\vec{r})}^2
    -
    \rho_0
    \abs{\vec{v}(\vec{r})}^2
    \right] \tensor{I}
    +
    \rho_0 \,
    \vec{v}^* \otimes \vec{v},
\end{equation}
where $\tensor{I}$ is the identity tensor
and where
$\kappa_0 = (\rho_0 c_0^2)^{-1}$ is the isentropic compressibility
of the medium given its density and speed of sound, $c_0$.
The first term on the right-hand side of
Eq.~\eqref{eq:stresstensor}
accounts for the Lagrangian energy density of the sound.
The second is the Reynolds stress
\cite{westervelt1951theory,yosioka1955acoustic}.
Integrating the normal component of the stress
over the surface, $S_i$, of the $i$-th particle
yields the time-averaged force experienced
by that particle:
\begin{equation}
    \label{eq:force}
    \vec{F}_i(\vec{r}_i)
    =
    -\frac{1}{2}\real{ \oiint_{S_i}
    \tensor{\sigma}(\vec{r}) \cdot \hat{n} \, d^2 r },
\end{equation}
\end{subequations}
where $\hat{n}(\vec{r})$ is the unit normal
to the particle's surface, and where
$S_i$ is referenced to the particle's position, $\vec{r}_i$.
In practice, $\vec{F}_i(\vec{r}_i)$ is most conveniently
obtained by setting
$p(\vec{r}) = \Pi_i(\vec{r})$ in Eq.~\eqref{eq:meatgrinder},
where $\Pi_i(\vec{r})$
is the pressure inside the $i$-th particle.
We obtain an expression for this interior field
by matching boundary conditions in a multipole expansion.

Referring to Fig.~\ref{fig:geometry},
the pressure wave incident on
particle $i$
can be expressed as
\begin{equation}
    \label{eq:incidentpressure1}
    p_0(\vec{s}_i)
    =
    p_0 \sum_{\ell=0}^\infty
    \sum_{m=-\ell}^\ell
    a_{\ell m}(k \vec{r}_i) \,
    j_\ell (k s_i) \,
    Y_\ell^m(\theta_i, \phi) ,
\end{equation}
in a spherical coordinate system,
$\vec{s}_i = (s_i, \theta_i, \phi) = \vec{r} - \vec{r}_i$,
centered on the particle
and oriented along the pressure wave's
wave vector, $\hat{k}$.
Distances in Eq.~\eqref{eq:incidentpressure1}
are scaled by the wave number in the medium,
$k = \omega/c_0$.
The incident wave's geometry is
expressed in terms of spherical
Bessel functions of the first kind, $j_\ell(kr)$, and
spherical harmonics, $Y_\ell^m(\theta, \phi)$.
Its structure is encoded
in the beam shape coefficients, $a_{\ell m} (k\vec{r}_i)$,
computed in the particle's frame of reference.

The wave scattered
by particle $i$ similarly
can be expressed as a multipole expansion \cite{zheng1995acoustic},
\begin{equation}
    \label{eq:scatteredpressure1}
    p_i(\vec{s}_i)
    =
    p_0 \sum_{\ell = 0}^\infty
    \sum_{m = -\ell}^\ell
    b_{\ell m}^{(i)} (k \vec{r}_i) \,
    h^{(1)}_\ell (k s_i) \,
    Y_\ell^m(\theta_i, \phi) ,
\end{equation}
in terms of 
spherical Hankel functions of the first kind, 
$h^{(1)}_\ell (kr)$.
The scattered wave's beam shape coefficients,
\begin{equation}
    b_{\ell m}^{(i)}(k \vec{r}_i)
    = 
    a_{\ell m} (k \vec{r}_i) \,
    B_{\ell m}^{(i)} ,
\end{equation}
are obtained from the incident wave's beam shape
coefficients by applying the particle's
scattering coefficients, $B_{\ell m}^{(i)}$.
These, in turn,
are obtained by requiring the
pressure and the normal component of the
velocity to be continuous at the particle's surface.
The same boundary conditions also yield
the transmission coefficients, $D_{\ell m}^{(i)}$, that
establish the interior pressure,
\begin{equation}
\label{eq:Pi}
    \Pi_i(\vec{s}_i)
    =
    p_0 \sum_{\ell = 0}^\infty
    \sum_{m = -\ell}^\ell
    d_{\ell m}^{(i)} (k \vec{r}_i) \,
    j_\ell(k_i s_i) \,
    Y_\ell^m(\theta_i, \phi),
\end{equation}
where the interior beam shape coefficients are
\begin{equation}
    d_{\ell m}^{(i)}(k \vec{r}_i)
    =
    a_{\ell m} (k \vec{r}_i) \,
    D_{\ell m}^{(i)}.
\end{equation}
Distances within the particle are scaled by
$k_i = \omega / c_i$, where
$c_i$ is the interior speed of sound.

\subsection{Scattering by spheres}
\label{sec:spheres}

For simplicity and clarity, we consider the
special case in which the particles are spheres,
each with its own radius, $a_i$, 
density, $\rho_i$, and interior speed of sound, $c_i$.
Continuity of the pressure at the $i$-th sphere's surface
requires
\begin{subequations}
    \label{eq:boundaryconditions}
\begin{equation}
    \left. p_0(\vec{s}_i)
    +
    p_i(\vec{s}_i) \right\vert_{s_i = a_i}
    =
    \left. \Pi_i(\vec{s}_i) \right\vert_{s_i = a_i}.
\end{equation}
Continuity of the normal component
of the velocity requires
\begin{equation}
    \rho_0 \left. \frac{\partial}{\partial s_i}
    \left[p_0(\vec{s}_i)
    +
    p_i(\vec{s}_i)\right]
    \right\vert_{s_i = a_i}
    =
    \rho_i \left. \frac{\partial}{\partial s_i}
    \Pi_i(\vec{s}_i)
    \right\vert_{s_i = a_i}.
\end{equation}
\end{subequations}
In agreement with previous studies
\cite{doinikov2001acoustic},
we find these boundary conditions are satisfied by
the scattering and transmission coefficients,
\begin{align}
    \label{eq:Blm}
    B_{\ell m}^{(i)}
    & =
     \frac{\lambda_i \, j_\ell(k a_i) \, j_\ell'(k_i a_i)
    -
    j_\ell'(k a_i) \, 
    j_\ell(k_i a_i)}{
    {h^{(1)}_\ell}'(k a_i) \, 
    j_\ell(k_i a_i)
    -
    \lambda_i \, h^{(1)}_\ell(k a_i) \, 
    j_\ell'(k_i a_i)}, \\
    \label{eq:Dlm}
    D_{\ell m}^{(i)}
    & =
    \frac{\rho_0}{\rho_i}
    \frac{j_\ell(k a_i) \, {h^{(1)}_\ell}'(k a_i)
    -
    j_\ell'(k a_i) \, h^{(1)}_\ell(k a_i)}{
    {h^{(1)}_\ell}'(k a_i) \, 
    j_\ell(k_i a_i)
    -
    \lambda_i \, 
    h^{(1)}_\ell(ka_i) \, 
    j_\ell'(k_i a_i)} ,
\end{align}
respectively, where
primes denote derivatives with respect
to arguments and where
$\lambda_i = \rho_0 c_0 / (\rho_i c_i)$
is the specific acoustic impedance
of the particle relative to that of the medium.

\subsection{The force on a sphere}
\label{sec:sphereforce}

Substituting Eq.~\eqref{eq:Dlm}
into Eq.~\eqref{eq:Pi} yields the pressure
within the $i$-th sphere.
The force on that sphere then follows from Eq.~\eqref{eq:meatgrinder},
\begin{equation}
    \vec{F}_i(\vec{r}_i)
    =
    F_0 \, \real{
    \sum_{\ell = 0}^\infty
    \sum_{m = -\ell}^\ell
    J_{\ell m}^{(i)} \,
    d_{\ell m}^{(i)} \,
    d_{\ell+1, m}^{(i)\ast}} \,
    \hat{k} .
\label{eq:waveforce}
\end{equation}
The magnitude of the force is set by a prefactor,
\begin{equation}
    \label{eq:forcescale}
    F_0 = \frac{p_0^2}{\rho_0 \, \omega^2},
\end{equation}
that depends on properties of the sound wave
in the medium.
Coupling between multipole moments
mediated by scattering at the sphere's surface
is described by the coefficients,
\begin{widetext}
\begin{multline}
\label{eq:Jlm}
    J_{\ell m}^{(i)}
    =
    \frac{1}{2} \left(\frac{\rho_i}{\rho_0}\right)^2
    \sqrt{
    \frac{(\ell - m + 1) (\ell + m + 1)}{
    (2\ell + 3) (2\ell + 1)}} \,
    \Bigg\{
    [m^2 - (k a_i)^2] \,
    j_\ell(x_i) \, j_{\ell+1}(x_i) \\
    + 
    \frac{\rho_0}{\rho_i} x_i
    \Big[
    \ell \, 
    j_\ell(x_i) \, j'_{\ell+1}(x_i)
    -
    (\ell+2) \,
    j'_\ell(x_i) \, j_{\ell+1}(x_i)
    \Big]
    -
    \left(\frac{\rho_0}{\rho_i} x_i\right)^2
    j'_\ell(x_i) \, j'_{\ell+1}(x_i)
    \Bigg\} ,
\end{multline}
\end{widetext}
where $x_i = k_i a_i$. 
Equation~\eqref{eq:Jlm} differs from
previously reported expressions 
\cite{yosioka1955acoustic,zheng1995acoustic,sapozhnikov2013radiation}
for the
coupling coefficients, $J_{\ell m}^{(i)}$,
which only include terms with $m = 0$.
The additional terms in the compete expression
are required for waves that lack azimuthal symmetry,
including the scattered waves
exchanged by pairs of particles.
Equation~\eqref{eq:Jlm} 
is analogous to Eq.~(2.32)
in Ref.~[\onlinecite{doinikov2001acoustic}],
but projects the force along $\hat{k}$
rather than $\hat{z}$, which is
more useful for computing interactions.

\subsection{A sphere in a standing wave}
\label{sec:standingwave}

As an illustrative example, we use this formalism to evaluate the force
exerted on the
$i$-th sphere by a plane standing wave,
\begin{equation}
\label{eq:planepressurewave}
    p_0(\vec{r})
    =
    p_0 \sin(k z),
\end{equation}
whose axis is aligned in the vertical direction, $\hat{z}$.
The incident field's beam shape coefficients
\cite{doinikov1994sphere,zheng1995acoustic},
\begin{align}
    \label{eq:beamshapecoeffs}
    a_{\ell m}^{(0)}(k \vec{r}_i)
    & =
    4 \pi (-1)^{\ell - m}
    \sin\left(k z_i - \ell \frac{\pi}{2} \right)
    Y_\ell^{-m}(\alpha, 0) ,
\end{align}
depend on the particle's height, $z_i$, above
the nodal plane at $z = 0$.
Referring to the coordinate system from Fig.~\ref{fig:geometry},
the $\hat{z}$-oriented incident wave
has $\alpha = 0$.
In the absence of other particles,
we can use Eq.~\eqref{eq:waveforce} to compute the force on particle $i$ due to the
incident field:
\begin{equation}
   \label{eq:Fj0}
   \vec{F}_{i}(\vec{r}_i) =
    \frac{\pi}{3} F_0 \, (k a_i)^3
    \left(\frac{\kappa_i}{\kappa_0}
    -
    \frac{3\rho_i}{\rho_0+2\rho_i}\right)
    \sin(2k z_i) \, \hat{z}.
\end{equation}
Equation~\eqref{eq:Fj0} includes only terms
at monopole order ($\ell = 0$) in the multipole expansion,
and agrees with the standard
Gor'kov expression \cite{doinikov2001acoustic}
for the leading-order acoustic trapping force.
Contributions from higher multipole terms, 
$\ell \geq 1$, appear
at order $(k a_j)^5$ in the dimensionless particle size
and therefore can be neglected for spheres that are smaller than the wavelength of sound, $k a_j < 1$.

The Gor'kov force vanishes at both nodes and antinodes
of the incident pressure wave.
The prefactor of $\vec{F}_{i}$ is negative for
dense spheres ($\rho_i < \rho_0$ and $\kappa_i > \kappa_0$).
Such particles are stably trapped at pressure nodes.
Bubbles, by contrast, are stably trapped at antinodes.
This distinction qualitatively differentiates
the acoustic force landscape experienced by
dense spheres and bubbles.
It also establishes qualitatively different contexts
for their wave-mediated interactions.

\section{Sound-Mediated Pair Interactions}
\label{sec:interactions}

\subsection{Acoustic forces on a pair of particles}

As illustrated schematically in Fig.~\ref{fig:geometry},
the wave scattered by particle $j$
interferes with the external wave
incident on particle $i$,
\begin{equation}
    p(\vec{r})
    =
    p_0(\vec{r}) + p_j(\vec{r} - \vec{r}_j) ,
\end{equation}
and therefore contributes
to the force experienced by particle $i$.
In principle, the second particle scatters
a portion of $p(\vec{r})$ back to the first,
giving rise to a hierarchy of exchanged waves.
For simplicity, we invoke the first
Born approximation and consider only
the first exchange of scattered waves.

Computing the force on particle $i$
requires an expression for the
interior pressure, $\Pi_i(\vec{r})$,
and thus an expression for the
first particle's scattered wave,
$p_j(\vec{r} - \vec{r}_j)$,
in spherical coordinates, $\vec{s}_i$
centered on $\vec{r}_i$.
To facilitate the projection, we align the axis
of the coordinate system along
$\vec{r}_{ij} = \vec{r}_i - \vec{r}_j$,
as shown in Fig.~\ref{fig:geometry},
and set the angle $\alpha$ in Eq.~\eqref{eq:beamshapecoeffs} accordingly.
In this coordinate system, the
pressure wave scattered by particle $j$,
\begin{equation}
    \label{eq:coordinatetransformation}
    p_j(\vec{s}_i)
    =
    p_0 \sum_{\ell = 0}^\infty
    \sum_{m = -\ell}^\ell
    a^{(ij)}_{\ell m}(k \vec{s}_i) \,
    j_\ell(k s_i) \, Y_\ell^m(\theta_i, \phi),
\end{equation}
is incident on particle $i$ with beam shape coefficients,
\begin{subequations}
\label{eq:interactioncoeffs}
\begin{equation}
    a^{(ij)}_{\ell m}
    =
    \sum_{n = 0}^\infty
    K_{n \ell m}(k r_{ij}) \,
    b^{(j)}_{n m}(k \vec{r}_j),
\end{equation}
that are projected from the scattered wave's coefficients 
with a projection kernel
\cite{gabrielli2001sphere},
\begin{equation}
\label{eq:projectioncoefficient}
    K_{n \ell m} (kr_{ij})
    =
    \sum_{s=0}^{\ell + n}
    \sqrt{\frac{2s+1}{4\pi}} \,
    C(n m \vert s0 \vert \ell m) \,
    h^{(1)}_s(kr_{ij}),
\end{equation}
that accounts for the particles' separation, 
$r_{ij} = \abs{\vec{r}_{ij}}$.
The projection coefficients
are expressed in terms of Wigner 3-j symbols
through
\begin{multline}
    C(n m|s0|\ell m)
    =
    i^{-n+s+\ell} (-1)^m \\
    \times \sqrt{4\pi(2\ell+1)(2s+1)(2n +1)} \\
    \times \begin{pmatrix}
    n & s & \ell \\
    0 & 0 & 0 \\
    \end{pmatrix}
    \begin{pmatrix}
    n & s & \ell \\
    -m & 0 & m \\
    \end{pmatrix}.
\end{multline}
The upper limit of the
sum in Eq.~\eqref{eq:projectioncoefficient}
reflects selection rules for the Wigner 3-j symbols \cite{edmonds1996quantum}.
\end{subequations}

Both the incident and scattered waves contribute to
the beam shape coefficients for the wave
inside particle $i$,
\begin{equation}
    d_{\ell m}^{(ij)}
    =
    \left(
    a_{\ell m}^{(0)}
    +
    a^{(ij)}_{\ell m}
    \right) D_{\ell m}^{(i)}.
\end{equation}
The particle's size and properties influence
the interior wave through the
transmission coefficients
from Eq.~\eqref{eq:Dlm}.
The net wave-mediated 
force on particle $i$ then follows from Eq.~\eqref{eq:waveforce}.
The complementary force on the neighboring particle
is obtained by exchanging labels $i$ and $j$.

The total force experienced by particle $i$ includes
contributions from the incident wave, the
scattered wave, and their interference.
To clarify the nature of the interparticle
interaction,
we assume that both particles are stably trapped
in the same nodal plane of the planar standing
wave described by Eq.~\eqref{eq:planepressurewave}.
The primary Gor'kov force therefore vanishes identically
and $a_{\ell m}^{(0)}(k\vec{r})$
from Eq.~\eqref{eq:beamshapecoeffs}
is evaluated at $z = 0$ and with $\alpha = \pi/2$
for both particles.
The acoustic force on particle $i$ therefore can
be attributed entirely to its wave-mediated interaction
with particle $j$,
\begin{equation}
    \label{eq:Fij}
    \vec{F}_{ij}(\vec{r}_{ij})
    =
    F_0 \,
    \real{
    \sum_{\ell = 0}^\infty
    \sum_{m = -\ell}^\ell
    J_{\ell m}^{(i)} \,
    d_{\ell m}^{(ij)} \,
    d_{\ell+1, m}^{(ij)\ast}} \,
    \hat{r}_{ij} .
\end{equation}
For this much-studied model system
\cite{konig1893hydrodynamisch,bjerknes1906fields,yosioka1955acoustic,garcia2014experimental,sapozhnikov2013radiation,silva2014acoustic},
the wave-mediated force depends
on the particles' separation
through the beam-shape coefficients,
$d_{\ell m}^{(ij)}$.

The pair interaction in this geometry
historically has been dubbed
the
K\"onig force for dense spheres
\cite{konig1893hydrodynamisch}
and
the secondary Bjerknes interaction
for bubbles
\cite{bjerknes1906fields}.
For simplicity, we formulate this force
in the Rayleigh approximation,
$ka_i, ka_j < 1$, so that we may reasonably
truncate the multipole expansion at
quadrupole order, $\ell = 2$.

\subsection{K\"onig interaction: Dense spheres in a standing plane wave}
\label{sec:particles}

Spheres that are denser than the fluid medium will be
localized in one of the
nodal planes of the standing wave.
The leading-order expression for the wave-mediated
force on particle $i$ due to particle $j$,
\begin{subequations}
\label{eq:particles}
\begin{equation}
    \vec{F}_{ij}^K(r)
    =
    - 2\pi \, F_0 \,
    \Phi_K(kr) \,
    \eta_i \eta_j \,
    \hat{r} ,
\end{equation}
is the K\"onig interaction \cite{konig1893hydrodynamisch}.
Its dependence on the particles separation,
\begin{equation}
    \Phi_K(kr)
    =
    \frac{
    \left[1 - \frac{1}{3} (kr)^2\right]
    \cos(kr)
    +
    kr \sin(kr)}{
    (kr)^4} ,
\end{equation}
shows that the wave-mediated interaction is
attractive when the particles are near contact
and changes sign at larger separations.
The particles' coupling constants
\cite{settnes2012forces},
\begin{equation}
    \eta_i 
    = 
    \frac{\rho_0 - \rho_i}{\rho_0 + 2 \rho_i}
    (k a_i)^3
\end{equation}
\end{subequations}
depend on their density mismatch with the
medium, but not on their compressibilities.
This is reasonable because dense particles
principally scatter the dipolar velocity field.

Equation~\eqref{eq:particles} reduces to 
the previously published form
for the pair interaction \cite{silva2014acoustic} in the special case of identical particles.
The asymmetric case appears not to have been reported previously.

Most previous studies of sound-mediated
pair interactions go no further than the
dipole approximation and conclude that
wave-mediated interactions are reciprocal
in general
\cite{zheng1995acoustic, doinikov2001acoustic}.
References~\cite{silva2014acoustic} and ~\cite{stclair2023dynamics} note the existence of nonreciprocal acoustic interactions between
pairs of bubbles
but suggest they are too small to be significant.
In fact, the nature of the pair interaction changes
qualitatively when higher-order multipole
contributions are taken into account.
These changes are present even in the 
first-scattering approximation, and take a surprisingly elegant form.

The leading quadrupole-order correction
to the K\"onig force,
\begin{equation}
\label{eq:F21densespheres}
    \vec{F}_{ij}(r)
    =
    \vec{F}^K_{ij}(r) \,
    (1 + \chi^K_{ij}) ,
\end{equation}
breaks the reciprocity of the
pair interaction with
a term,
\begin{subequations}
\label{eq:chiKij}
\begin{equation}
    \chi^K_{ij}
    =
    \alpha^K_{ij}
    +
    \beta^K_{ij} \, (k a_i)^2
    +
    \gamma^K_{ij} \, (k a_j)^2,
\end{equation}
that identifies roles for the spheres'
sizes and compositions through the
coefficients,
\begin{align}
    \label{eq:alphaKij}
    \alpha^K_{ij} &= -\frac{2}{3}\frac{\rho_0}{\rho_i}\\
    \beta^K_{ij}
    & = 
    - \frac{3}{10} 
    \left(
    1 - \frac{\kappa_j}{\kappa_0}+\frac{3}{2} \frac{\rho_0}{\rho_j}-\frac{2}{3}\frac{\rho_0}{\rho_i}
    \right)
    \\
    \gamma^K_{ij}
    & =
    -\frac{19}{18} +\frac{127}{210}\frac{\kappa_i}{\kappa_0} - \frac{67}{108}\frac{\rho_0}{\rho_i} .
\end{align}
\end{subequations}

Equation~\eqref{eq:chiKij} establishes
the conditions under which acoustically
levitated spheres experience nonreciprocal
interactions.
Because $\alpha_{ij}^K \neq \alpha_{ji}^K$,
spheres with different densities interact
nonreciprocally even if they have the same
size.
The different functional forms of
$\beta_{ij}^K$ and $\gamma_{ij}^K$
have the consequence that
spheres of different sizes
interact nonreciprocally even if they
are made of the same material.
Equation~\eqref{eq:chiKij} further
reveals that nonreciprocal effects
should be most
prominent for particles that are
comparable in size to the wavelength of sound,
$k a_i \lesssim 1$, and are 
nearly density-matched
to the medium, $\rho_i \gtrsim \rho_0$.

\subsection{Bjerknes interaction: Bubbles in a standing plane wave}
\label{sec:bubbles}

Bubbles are less dense than the medium
($\rho_i < \rho_0$)
and more compressible
($\kappa_i > \kappa_0$)
and so are localized
at antinodes of the pressure wave.
The wave-mediated force between acoustically
levitated bubbles is commonly known as the
secondary Bjerknes interaction \cite{bjerknes1906fields}
and broadly
resembles the K\"onig interaction between dense spheres 
from Eq.~\eqref{eq:particles}.
To leading nontrivial order in the small parameters,
$ka_i$, $\rho_i/\rho_0$ and $\kappa_0/\kappa_i$,
Eq.~\eqref{eq:Fij}
predicts that
the wave-mediated interaction between
two bubbles separated by distance
$r$ in an antinodal plane of a
standing wave is
\begin{subequations}
\label{eq:bjerknes}
\begin{equation}
    \vec{F}_{ij}^B(r)
    =
    -\frac{2\pi}{9} \,
    F_0 \,
    \Phi_B(kr) \,
    \eta_i \eta_j \,
    \hat{r},
\end{equation}
with coupling constants of the form \cite{settnes2012forces}
\begin{equation}
    \eta_i
    =
    \left(1 - \frac{\kappa_i}{\kappa_0} \right)
    (k a_i)^3 .
\end{equation}
The Bjerknes interaction is longer-ranged
than the K\"onig interaction,
\begin{equation}
    \Phi_B(kr)
    =
    \frac{\cos(kr) + kr \, \sin(kr)}{(k r)^2},
\end{equation}
\end{subequations}
and is attractive at short ranges.
Bubbles' coupling constants depend on
their compressibilities rather than their densities because bubbles
principally scatter the pressure field,
with a leading contribution from
monopole scattering.
Equation~\eqref{eq:bjerknes}
includes terms up to dipole
order ($\ell = 1$) in the multipole expansion
and agrees with previously reported expressions
\cite{doinikov2001acoustic,sapozhnikov2013radiation}
for this interaction at the same level of approximation.
Reference~[\onlinecite{yosioka1955acoustic}]
proposes a different numerical prefactor
because its derivation
imposes axisymmetry on the pressure field,
which is not appropriate for multipole contributions
with $m \neq 0$.
As for the K\"onig interaction between
dense spheres, the leading-order 
Bjerknes interaction
is reciprocal, $\vec{F}^B_{21}(r) = - \vec{F}^B_{12}(r)$, 
even for bubbles of different sizes
and compositions.

The quadrupole-order expression,
\begin{equation}
    \vec{F}_{ij}(r)
    =
    \vec{F}_{ij}^B(r) \, \left(1 + \chi^B_{ij}\right)
\end{equation}
introduces a correction,
\begin{subequations}
\label{eq:chiBij}
\begin{equation}
    \chi^B_{ij}
    =
    \alpha^B_{ij}
    +
    \beta^B_{ij} \, (k a_j)^2
    +
    \gamma^B_{ij} \, (k a_i)^2
\end{equation}
that depends on the bubbles' compositions
and sizes.
Additional terms and higher
multipole contributions all appear
at $\order{(ka_j)^3}$ or $\order{(ka_i)^3}$
and so can be neglected in the Rayleigh
approximation.
Expressing the coefficients to leading
nontrivial order yields
\begin{align}
    \label{eq:alphaBij}
    \alpha^B_{ij}
    & =
    -\frac{3}{2}
    \frac{\kappa_0}{\kappa_i}
    \frac{\rho_i^2}{\rho_0^2}
    \\
    \beta^B_{ij}
    & =
    - \frac{2}{3}
    +
    \frac{1}{3}
    \frac{\kappa_j}{\kappa_0}
    \left( 
    1 + \frac{1}{5} \frac{\rho_j}{\rho_0}
    \right)
    \\
    \gamma^B_{ij}
    & =
    - \frac{2}{3}
    +
    \frac{1}{3} \frac{\kappa_i}{\kappa_0}
    \left(
    1
    +
    \frac{1}{5}
    \frac{\rho_i}{\rho_0}
    \right) ,
\end{align}
\end{subequations}
from which we conclude the wave-mediated interactions
between pairs of bubbles are
reciprocal unless the bubbles are
composed of different materials.
Size-dependent nonreciprocity emerges at
higher orders in the small parameters,
$\rho_i/\rho_0$, $\rho_j/\rho_0$,
$\kappa_0/\kappa_i$ and $\kappa_0/\kappa_j$.
Comparing Eq.~\eqref{eq:alphaBij} with
Eq.~\eqref{eq:alphaKij} suggests that
$\alpha^B_{ij} \ll \alpha^K_{ij}$, which means
that composition-dependent nonreciprocity should be significantly weaker for bubbles than
for dense particles.
This is consistent with earlier reports
\cite{silva2014acoustic,stclair2023dynamics}.
Nonreciprocal effects 
should be most evident in systems such
as emulsions where the ``bubbles'' are
nearly density matched to the medium
and have comparable compressibility.
The nature of the pairwise nonreciprocity reported in Eq.~\eqref{eq:chiKij}
and Eq.~\eqref{eq:chiBij}
complements a recent report
of nonreciprocal wave-mediated interactions
that arise from the viscosity of the medium
thanks to streaming effects that we do not consider
\cite{doinikov2023self}.

\section{Light-mediated pair interactions}
\label{sec:opticalbinding}

The electric field acting on 
a dielectric particle
at position $\vec{r}_i$ is the
superposition of the incident field,
$\vec{E}_0(\vec{r})$, and the field
scattered by its neighbor at $\vec{r}_j$:
\begin{equation}
    \vec{E}_i(\vec{r}_i) 
    = 
    \vec{E}_0(\vec{r}_i)
    +
    \alpha_j \, \vec{E}_j(\vec{r}_j) \,
    \tensor{G}(\vec{r}_{ij}),
\end{equation}
where $\alpha_j$ is the scattering
coefficient for particle $j$,
$\vec{E}_j(\vec{r}_j)$ is the total field
incident on particle $j$ and
$\vec{r}_{ij} = \vec{r}_i - \vec{r}_j$ is
the interparticle separation.
The tensorial Green function \cite{dapasse1994optical},
\begin{equation}
    \tensor{G}(\vec{r}) 
    = 
    [\lambda(r) - \mu(r)] \, 
    \hat{\vec{r}} \otimes \hat{\vec{r}} 
    +
    \mu(r) \, \tensor{I},
\end{equation}
expresses propagation of the components
of the scattered wave
from $j$ to $i$ 
in terms of the partial waves
\begin{equation}
    \lambda(r) = \frac{e^{ikr}}{r^3}(1 - ikr )
\end{equation}
and 
\begin{equation}
    \mu(r) = \frac{e^{ikr}}{r^3}[(kr)^2 - 1 + ikr].
\end{equation}
For simplicity, we assume that the incident
field is uniform, $\vec{E}_0(\vec{r}) = \vec{E}_0$.
When the incident field is linearly polarized either
transverse or parallel
to $\vec{r}_{ij}$, the magnitude of the
field incident on
particle $i$ can be expressed 
self-consistently as
\begin{equation}
    E_i(r) 
    = 
    E_0 \,
    \frac{1 + \alpha_j \, G_{ij}}{
    1 - \alpha_i \alpha_j \, G_{ij}^2(\vec{r}_{ij})},
\end{equation}
where $G_{ij}$ is the transverse
or longitudinal
component of $\tensor{G}(\vec{r}_{ij})$, respectively.

The total force on particle $i$ due to particle $j$
follows from the field \cite{chaumet2000time}:
\begin{equation}
    \vec{F}_{i}(\vec{r}_i) 
    = 
    \frac{1}{2} \sum_{n=1}^3
    \real{\alpha_i^\ast \,
    E_{i,n}^\ast 
    \frac{\partial E_{i,n}}{\partial r} }
    \hat{e}_n,
\end{equation}
where the subscript, $n$, refers to the Cartesian
coordinates.
This expression includes the pair interaction,
\begin{widetext}
\begin{equation}
    \vec{F}_{ij}(\vec{r}_{ij}) 
    = 
    \frac{1}{2} E_0^2
    \real{
    \alpha_i^\ast \alpha_j
    \left(
    1 
    +
    2 \alpha_i G_{ij} 
    + \alpha_j^\ast G_{ij}^\ast
    \right)
    \frac{\partial G_{ij}}{\partial r}
    } \hat{r}_{ij}
    +
    \order{\alpha_i \alpha_j G_{ij}^2},
\end{equation}
which we express in the limit of weak
scattering, $\abs{\alpha_j G_{ij}} \ll 1$.
Specializing to the case of transverse polarization,
$G_{ij} = \lambda(r_{ij})$, 
 we obtain an expression for the pair interaction,
\begin{equation}
    \vec{F}_{ij}(\vec{r}_{ij}) 
    = 
    - \frac{3}{2} E_0^2 \,
    \real{
    \alpha_i^\ast \alpha_j
    \left[
    -\frac{(kr_{ij})^2}{3}
    +
    i k r_{ij} + 1
    \right]
    e^{ikr_{ij}}
    } \hat{r}
    +
    \order{k^6\alpha_i^3}.
\end{equation}
\end{widetext}
Factoring the separation dependence into $\Phi(kr)$ yields the expression for the
optical binding force presented in Eq.~\eqref{eq:opticalbinding}.
The reciprocal part of the interaction
is obtained by noting 
that the product of particle polarizabilities can be written as
\begin{equation}
    \alpha_i^\ast \alpha_j 
    = 
    \alpha_i'\alpha_j' + \alpha_i''\alpha_j''
    + i ( \alpha_i'\alpha_j'' - \alpha_i''\alpha_j')
\end{equation}
and taking the real part.
The nonreciprocal component 
comes from the imaginary part.
We can then write the optical binding force in the form of Eq.~\eqref{eq:decomposedopticalforce}.
The final expression is obtained by applying
Eq.~\eqref{eq:polarizability} and Eq.~\eqref{eq:clausiusmossotti} and neglecting terms of higher order than $\order{k^6 a_i^6}$.

\section{Activity as a function of polydispersity}
\label{sec:polydispersity}

The steady-state velocity of a pair of acoustically levitated
particles arises from the competition between unbalanced center-of-mass
forces and viscous drag.
Hydrodynamic coupling reduces the drag on a pair
of spheres \cite{happel2012low} relative to
the sum of Stokes contributions, which means that
\begin{align}
    \vec{v}_{ij} 
    & =
    \frac{\chi_{ij} - \chi_{ji}}{6 \pi \eta_0 \, (a_i + a_j)} \,
    F^K_{ij}(a_i + a_j) \, \hat{r}_{ij} ,
\end{align}
is an underestimate for the
pair velocity due to emergent activity.
Equation~\eqref{eq:activity} and
Eq.~\eqref{eq:alphaKij} then suggest
that two dense spheres composed of the same material 
have an emergent activity that depends on their composition
and sizes as
\begin{equation}
    \label{eq:pairactivity}
   A_S(a_i, a_j) 
   =
   2 \pi^2 f_1^4 (k a_i)^7  (\beta - \gamma)^2 k\mu 
   \xi^6 \,
   \frac{(1 - \xi)^2}{(1 + \xi)^7} ,
\end{equation}
where
\begin{equation}
    f_1 = \frac{\rho_0 - \rho_i}{\rho_0 + 2 \rho_i},
\end{equation}
is the dipole polarizability of a sphere of density
$\rho_i$ and $\xi = a_j/a_i$ is the spheres' size ratio.
For spheres made of the same material, $\beta_{ij}=\beta_{ji}=\beta$ and $\gamma_{ij}=\gamma_{ji}=\gamma$.

We compute the ensemble-averaged pair
activity by assuming
that the spheres' radii are drawn from
$\Gamma(\mu^2/\sigma^2,\sigma^2/\mu)$ with mean radius $\mu$ and variance $\sigma^2$,
with probability density
\begin{equation}
    p(a)
    =
    \frac{1}{a} \left(\frac{\mu a}{\sigma^2}\right)^{\frac{\mu^2}{\sigma^2}}
    \frac{
    e^{-\frac{\mu a}{\sigma^2}}}{\Gamma\left(\frac{\mu^2}{\sigma^2}\right)}.
\end{equation}
With this assumption, $\xi$ is a
random variable independent of $a_j$ that is drawn from the Beta prime
distribution $\beta'(\mu^2/\sigma^2, \mu^2/\sigma^2, 1,1)$,
with probability density
\begin{equation}
    p(\xi) = \frac{\xi^{-1+\mu^2/\sigma^2}(1+\xi)^{-2\mu^2/\sigma^2}}{B(\mu^2/\sigma^2, \mu^2/\sigma^2)}.
\end{equation}
Averaging Eq.~\eqref{eq:pairactivity} over $a_i \in [0, \infty]$ and over $\xi \in [0,1]$ yields
the approximate mean pair activity,
\begin{subequations}
\label{eq:2particletheory}
\begin{multline}
    A_S(X)
    = 
    A_0 X^{14}
    \frac{\Gamma\left(6 + \frac{1}{X^2}\right) 
    \Gamma\left(7 + \frac{1}{X^2}\right)
    \Gamma\left(\frac{2}{X^2}\right)}{
    \Gamma\left(9 + \frac{1}{X^2}\right)
    \Gamma^3\left(\frac{1}{X^2}\right)
    } 
    \\
    \times
    \hypgeo{2}{1}\left(6 + \frac{1}{X^2}, 7 + \frac{2}{X^2}; 
    9 + \frac{1}{X^2}; -1 \right) ,
\end{multline}
as a function of the polydispersity, $X = \mu/\sigma$.
The activity is expressed in terms of the gamma function, $\Gamma(n)$,
and the Gauss hypergeometric function,
$\hypgeo{2}{1}(\alpha, \beta; \gamma; z)$.
The overall activity scale,
\begin{equation}
    A_0 
    = 
    2 \pi^2(\beta-\gamma)^2
    (k \mu)^8 f_1^4 ,
\end{equation}
\end{subequations}
depends strongly on the mean particle size, $\mu$,
relative to the wavelength.

We obtain the leading-order scaling of emergent activity as a function of polydispersity through a mean field approximation. 
We assume that the  radius of each particle
differs only slightly from the mean,
$a_i = \mu (1 + \epsilon_i)$, and 
that $\epsilon_i$ is drawn from a normal
distribution with width $\sigma/\mu$.
The emergent activity of two dense spheres is then
\begin{equation}
    \label{eq:pairactivitymeanfield}
   A_S(a_i, a_j) 
   =
   A_0
    \frac{(1+\epsilon_i)^6\,
   (1+\epsilon_j)^6\,
   (\epsilon_i-\epsilon_j)^2\,}{(2 + \epsilon_i+\epsilon_j)^7} .
\end{equation}
Applying the binomial expansion to
the denominator under the assumption that
$\epsilon_i + \epsilon_j \ll 2$, 
and averaging the resulting polynomial
expression over $\epsilon_i$ and $\epsilon_j$
then yields the leading order behavior
\begin{equation}
    A_S(X) = A_0(X^2+\order{X^4})
\end{equation}
that appears in Eq.~\eqref{eq:leadingactivity}.

\bibliography{nonreciprocity}
\end{document}